\documentclass[preprint,showpacs,preprintnumbers,amsmath,amssymb,aps]{revtex4}

\usepackage{graphicx}
\usepackage{dcolumn}
\usepackage{bm}

\begin{document}

\preprint{APS/123-QED}

\title{Various regimes of quantum behavior in $S$ = 1/2 Heisenberg antiferromagnetic chain with fourfold  periodicity}

\author{H. Yamaguchi$^1$, T. Okubo$^2$, K. Iwase$^1$, T. Ono$^1$, Y. Kono$^2$, \\S. Kittaka$^2$, T. Sakakibara$^2$, A. Matsuo$^2$, K. Kindo$^2$, and Y. Hosokoshi$^1$}
\affiliation{
$^1$Department of Physical Science, Osaka Prefecture University, Osaka 599-8531, Japan \\ 
$^2$Institute for Solid State Physics, the University of Tokyo, Chiba 277-8581, Japan \\
}

Second institution and/or address\\
This line break forced

\date{\today}

\begin{abstract}
We have succeeded in synthesizing single crystals of the verdazyl radical $\beta$-2,6-Cl$_2$-V [= $\beta$-3-(2,6-dichlorophenyl)-1,5-diphenylverdazyl]. 
The $ab$ $initio$ MO calculation indicated the formation of an $S=1/2$ Heisenberg antiferromagnetic chain with four-fold magnetic periodicity consisting of three types of exchange interactions.
We have successfully explained the magnetic and thermodynamic properties based on the expected spin model by using the quantum Monte Carlo method.
Furthermore, we revealed that the alternating and unique Ising ferromagnetic chains become effective in the specific field regions and observed a cooperative phenomenon caused by the magnetic order and quantum fluctuations.
These results demonstrate that verdazyl radical could form unconventional spin model with interesting quantum behavior and provide a new way to study a variety of quantum spin systems.
\end{abstract}

\maketitle
\section{INTRODUCTION}
Quantum spin systems have been investigated intensively over the last few decades, both experimentally and theoretically.
They provide unique many-body phenomena caused by their strong quantum fluctuations.
In previous studies, simple quantum spin systems consisting of one or two kinds of interactions have been fully understood. 
Among such systems, the $S$=1/2 Heisenberg antiferromagnetic chain (HAFC) is the simplest example, and its ground state is known to be described as a Tomonaga-Luttinger liquid (TLL), which is a quantum critical state with fermionic $S$ = 1/2 spinon excitations~\cite{tomonaga, luttinger}.
Haldane's conjecture in 1983~\cite{Haldane} stimulated studies on HAFC with higher spin values, and subsequent extensive investigations established the presence of the Haldane gap in HAFCs with integer spin values~\cite{NENP, MnS2}.
The $S$=1/2 antiferromagnetic (AF) spin-ladders have also attracted much attention as simple quantum spin systems in relation to quantum phase transitions and high-$T_{\rm{c}}$ superconductors~\cite{ladder_the,ladder_exp,SL1, SLCS}.  
In such gapped spin systems, the field-induced quantum phase transitions to the gapless TLL is the focus of intensive research towards the understanding of the quantum critical point~\cite{gap1,gap2}.
Although the ground states in those fundamental simple quantum spin systems have been fully understood, more complicated quantum spin systems with multiple exchange interactions have not been sufficiently investigated, owing to a lack of experimental realizations.
It is of great interest and importance to synthesize and investigate unconventional quantum systems to further our understanding of quantum effect in magnetic materials.

A large number of experiments on inorganic materials forming quantum spin systems have been reported previously.
However, most of them are characterized by large values for the exchange couplings, and these strong magnetic interactions make the observation of magnetic-field-dependent behavior in laboratory experiments difficult.
Elemental substitution or application of pressure is required to modulate the magnetic interactions in such inorganic compounds.
On the other hand, we can modify the magnetic interactions by using simple chemical modifications in organic radical compounds.  
We focused on a verdazyl radical with delocalized $\pi$-electron spins extending over the molecule. 
Owing to a lack of high-quality single crystals, most previous studies on verdazyl-based materials have not focused on their crystal structures, which are vital for gaining understanding of their magnetic states. 
Recently, we established synthetic techniques for preparing high-quality verdazyl-based single crystals and demonstrated modulations of the magnetic interactions by using simple chemical modifications~\cite{p-BIP-V2,3Cl4FV,2Cl6FV,mPhV2}.
This chemical modification method enabled us to synthesize varieties of unconventional magentic materials.

In this paper, we report the first model compound of an $S=1/2$ HAFC with a four-fold magnetic periodicity. 
We have succeeded in synthesizing the verdazyl radical $\beta$-2,6-Cl$_2$-V [= 3-(2,6-dichlorophenyl)-1,5-diphenylverdazyl] and solved its crystal structure. 
An $ab$ $initio$ molecular orbital (MO) calculation indicates the formation of an $S=1/2$ HAFC with a four-fold magnetic periodicity. 
We successfully explain the experimental results for magnetic and thermodynamic properties as the expected spin model by using the quantum Monte Carlo (QMC) method. 
From the perturbation treatment of the spin Hamiltonian, we found that an alternating and unique Ising ferromagnetic chains become effective in the specific field regions.
The field-induced partial ordering appears as a cooperative phenomenon of an effective long-range order and quantum spin fluctuations.

\section{EXPERIMENTAL}
The synthesis of 2,6-Cl$_2$-V, whose molecular structure is shown in Fig. 1(a), was mainly performed through a conventional procedure~\cite{procedure}.
The recrystallization in acetonitrile yielded crystals of $\alpha$ (green octahedron) and $\beta$ (green needle) phases, which are defined based on the values of unit cell volume per number of molecules ($V$/$Z$). 
X-ray intensity data were collected at 296 K using a Rigaku CCD Mercury diffractometer. 
The structure was solved by a direct method and was refined by full-matrix least squares techniques using the SHELX-97~\cite{SHELX-97}.
 
The magnetic susceptibility and magnetization curves were measured using a commercial SQUID magnetometer (MPMS-XL, Quantum Design) and a capacitive Faraday magnetometer with a dilution refrigerator.
The experimental results were corrected for the diamagnetic contribution of -3.07$\rm{\times}$10$^{-4}$ emu$\cdot$mol$^{-1}$, which is determined based on the QMC analysis to be described and close to that calculated by Pascal's method.
The specific heat was measured with a commercial calorimeter (PPMS, Quantum Design) by a thermal relaxation method above 2.0 K and by an adiabatic method down to 0.35 K. 
The high-field magnetization at pulsed magnetic fields of up to about 75 T was measured using a non-destructive pulse magnet at the Institute for Solid State Physics at the University of Tokyo. 
All experiments were performed using small randomly oriented single crystals with typical dimensions of 2.0$\rm{\times}$0.5$\rm{\times}$0.5 mm$^3$.

 The $ab$ $initio$ MO calculations for real molecules are feasible and form a powerful approach for investigating the electric properties of the molecules; these calculations were performed using the UB3LYP method as broken-symmetry (BS) hybrid density functional theory calculations. All the calculations were performed using the Gaussian 09 program package and 6-31G basis sets.
The convergence criterion was 10$^{-8}$ hartree. 
To estimation the intermolecular magnetic interaction, we applied our $J$ evaluation scheme to multispin systems that have been studied previously using the Ising approximation~\cite{MOcal}.

Our QMC code is based on the directed loop algorithm in the stochastic series expansion representation~\cite{QMC2}. The calculation was performed for a system of 256 spins under the periodic boundary condition.

\section{RESULTS}

\subsection{Crystal structure and magnetic model}
The crystallographic data are summarized in Table I~\cite{crystal}.
The verdazyl ring ( which includes four nitrogen atoms), the upper two phenyl rings, and the bottom 2,6-dichlorophenyl ring are labeled as ${\rm{R}_{1}}$, ${\rm{R}_{2}}$, ${\rm{R}_{3}}$, and ${\rm{R}_{4}}$, respectively, as shown in Fig. 1(a).
There are crystallographically independent molecules, M$_1$ and M$_2$, in the crystal, as shown in Fig. 1(b). 
It is significant that both molecules are no longer planar, and the dihedral angles of ${\rm{R}_{1}}$-${\rm{R}_{2}}$, ${\rm{R}_{1}}$-${\rm{R}_{3}}$, ${\rm{R}_{1}}$-${\rm{R}_{4}}$ for M$_1$ and M$_2$ are about 34$^{\circ}$, 38$^{\circ}$, 89$^{\circ}$ and 64$^{\circ}$, 33$^{\circ}$, 77$^{\circ}$, respectively. 
We performed $ab$ $initio$ MO calculations to evaluate quantitatively the dominant intermolecular magnetic interactions on all molecular pairs within 4.5 $\rm{\AA}$.
Consequently, we found that there are three types of dominating interactions $J_{\rm{1}}$, $J_{\rm{2}}$, and $J_{\rm{3}}$. 
They are evaluated to be $J_{\rm{1}}/k_{\rm{B}}$ = 5.6 K, $J_{\rm{2}}/k_{\rm{B}}$ = 18.7 K, and $J_{\rm{3}}/k_{\rm{B}}$ = 72.2 K, which are defined in the following eq. (1), and form an $S$ = 1/2 HAFC with a four-fold magnetic periodicity consisting of $J_{\rm{1}}$-$J_{\rm{2}}$-$J_{\rm{3}}$-$J_{\rm{2}}$ interactions along the $a$-axis, as shown in Fig. 1(c).
The MO calculation indicated that about 64\% of the total spin-density is present on ${\rm{R}_{1}}$ for both molecules. 
While ${\rm{R}_{2}}$ and ${\rm{R}_{3}}$ have about 15 \% $\sim$ 20 \% of the relatively large spin-densities for each phenyl ring, the ${\rm{R}_{4}}$ has a spin-density less than about 1.3 \% for both molecules. 
Therefore, the intermolecular interactions are mainly caused by short contacts of N or C related to the R$_1$$\sim$R$_3$ rings.
The evaluated interactions $J_{\rm{1}}$, $J_{\rm{2}}$, and $J_{\rm{3}}$ correspond to interactions between the M$_1$-M$_1$, M$_1$-M$_2$, and M$_2$-M$_2$ molecular pairs, respectively, as shwon in Fig. 1(b).  
The C-C short contact between the M$_1$ molecules related to $J_{\rm{1}}$, which is doubled by an inversion symmetry, has a longer distance of 3.70 $\rm{\AA}$ than the other two, and thus a relatively weak interaction exist there.
Although the C-C contact between M$_1$ and M$_2$ molecules related to $J_{\rm{2}}$ has a short distance of 3.40 $\rm{\AA}$, the small overlap of the $\pi$ orbitals, which expand perpendicular to the planes, due to a relatively large dihedral angle between the related planes should make the interaction weak. 
Conversely, the similarly C-C short contact of 3.40 $\rm{\AA}$ between the M$_2$ molecules related to $J_{\rm{3}}$, which is also doubled by an inversion symmetry, has a parallel stacking of the related planes, and thus the strongest interaction is expected there.
Additionally, a weak ferromagnetic interchain interaction of $J^{\rm{\prime}}/k_{\rm{B}}$ = -1.3 K was evaluated between the M$_1$ and M$_2$ molecules from the MO calculation.
Despite the N-C short contact of 3.54 $\rm{\AA}$, only a slight overlap of the $\pi$ orbitals can be expected because of its almost side-by-side contact, as shown in Fig. 1(d).
This interchain interaction forms square lattice perpendicular to the chain direction along the $a$ axis.

\begin{table}
\caption{Crystallographic data for $\beta$-2,6-Cl$_2$-V}
\label{t1}
\begin{center}
\begin{tabular}{cc}
\hline 
Formula & C$_{20}$H$_{15}$Cl$_{2}$N$_{4}$\\
Crystal system & Monoclinic \\
Space group & $P$2$_1$/$n$ \\
$a/\rm{\AA}$ & 17.307(6) \\
$b/\rm{\AA}$ & 11.662(4) \\
$c/\rm{\AA}$ & 18.464(7) \\
$\beta$/degrees & 95.968(7) \\
$V$/$\rm{\AA}^3$ & 3706(2) \\
$Z$ & 8 \\
$D_{\rm{calc}}$/g cm$^{-3}$ & 1.370 \\
Temperature & RT \\
Radiation & Mo K$\rm{\alpha}$ ($\lambda$ = 0.71075 $\rm{\AA}$  ) \\
Total reflections & 6469 \\
Reflection used & 4167 \\
Parameters refined & 469 \\
$R$ [$I>2\sigma(I)$] & 0.0670 \\
$R_w$ [$I>2\sigma(I)$] & 0.1668 \\
Goodness of fit & 1.052 \\
\hline
\end{tabular}
\end{center}
\end{table}

\begin{figure}[t]
\begin{center}
\includegraphics[width=18pc]{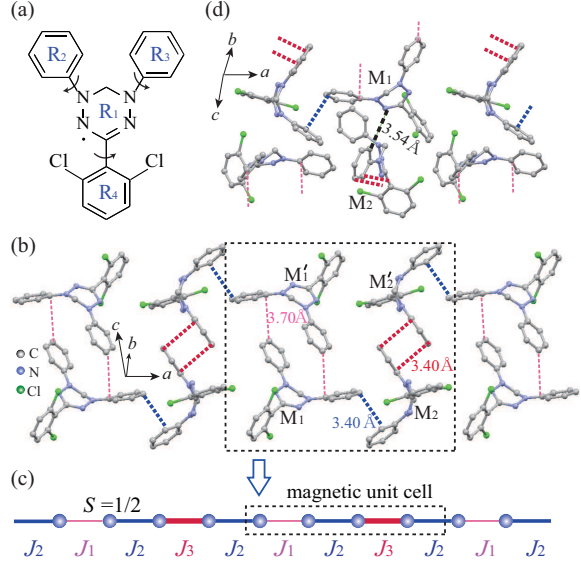}
\caption{(color online) (a) The molecular structure of $\beta$-2,6-Cl$_2$-V. (b) Crystal structure forming a chain along the $a$-axis, and (c) corresponding $S$ = 1/2 HAFC with four-fold magnetic periodicity. (d) Molecular packing associated with the interchain interaction $J^{\rm{\prime}}$. Broken lines indicate C-C and N-C short contacts. Hydrogen atoms are omitted for clarity.}\label{f1}
\end{center}
\end{figure}

\begin{figure}[t]
\begin{center}
\includegraphics[width=18pc]{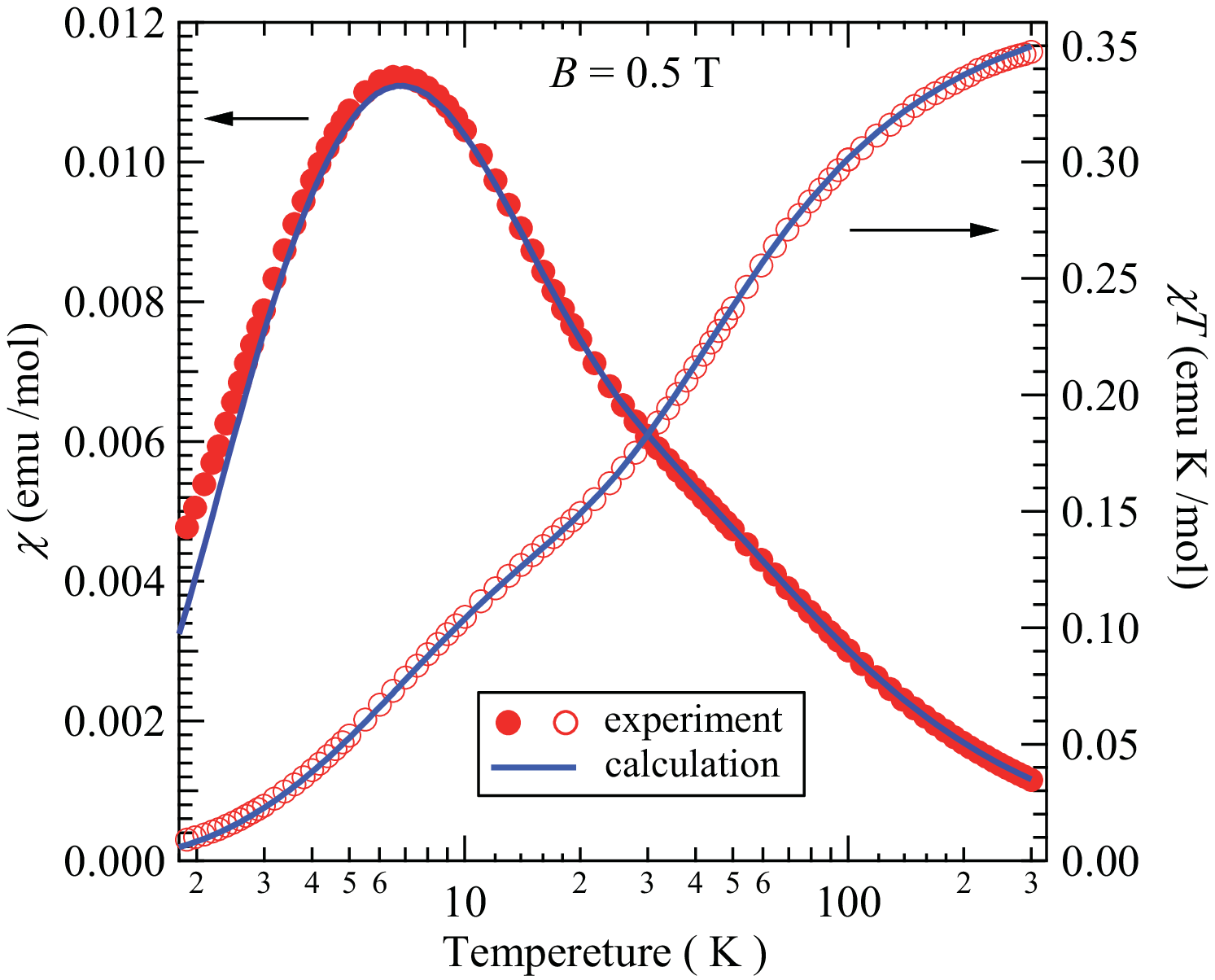}
\caption{(color online) Temperature dependence of magnetic susceptibility ($\chi$ = $M/H$) and ${\chi}T$ of $\beta$-2,6-Cl$_2$-V at 0.5 T. The solid lines represent the calculated results for the $S$ = 1/2 HAFC with four-fold periodicity.}\label{f2}
\end{center}
\end{figure}

\subsection{Magnetic susceptibility}
Figure 2 shows the temperature dependence of the magnetic susceptibility ($\chi$ = $M/H$) at 0.5 T.
Above 200 K, it follows the Curie-Weiss law, and the Weiss temperature is estimated to be ${\theta}_{\rm{W}}$ = -21.4(6) K. 
We observe a broad peak and shoulder at about 6.5 K and 30 K, respectively. 
Below 6.5 K, $\chi$ decreases with decreasing temperatures, which indicates the existence of a nonmagnetic ground state separated from the excited states by an energy gap.
The temperature dependence of ${\chi}T$ shows a gradual change at the shoulder.
The two-step decrease of ${\chi}T$ with decreasing temperatures is expected to be associated with formations of two types of singlet states originating from different types of exchange interactions.    
We calculated the magnetic susceptibility based on the $S$ = 1/2 HAFC with four-fold magnetic periodicity by using the QMC method. 
The spin Hamiltonian is expressed as
\begin{equation}
\mathcal {H} = J_{\rm{1}}{\sum^{}_{ij}}\textbf{{\textit S}}_{i}{\cdot}\textbf{{\textit S}}_{j}+J_{\rm{2}}{\sum^{}_{kl}}\textbf{{\textit S}}_{k}{\cdot}\textbf{{\textit S}}_{l}+J_{\rm{3}}{\sum^{}_{mn}}\textbf{{\textit S}}_{m}{\cdot}\textbf{{\textit S}}_{n}-g{\mu _B}H{\sum^{}_{i}}\textbf{{\textit S}}_{i},
\end{equation}
where $\textbf{{\textit S}}$ is the spin operator, $g$ the $g$-factor, $g$ = 2.00, ${\mu}$$_B$ the Bohr magneton, and $H$ the external magnetic field. 
We obtained good agreement between the experiment and calculation by considering the parameters $J_{\rm{1}}/k_{\rm{B}}$ = 8.6 K, $J_{\rm{2}}/k_{\rm{B}}$ = 34 K, and $J_{\rm{3}}/k_{\rm{B}}$ = 77 K, as shown in Fig. 2. 
Comparing the experimental results and the $ab$ $initio$ values, these parameters can be regarded as consistent with those evaluated from the MO calculation.

\subsection{Magnetization curve}
Figure 3 shows the magnetization curves at various temperatures.
We found a wide range 1/2 magnetization plateau from 14 T to 70 T at 1.3 K.
The magnetization curves at lower temperatures exhibit a zero-field excitation gap of about 3.5 T, as shown in the lower inset of Fig. 3.
The field derivative of the magnetization curve ($dM/dH$) indicates an asymmetric double peak structure associated with the phase boundaries, as shown in the upper inset of Fig. 3.
Our QMC calculations at 0.60 and 1.3 K, using the parameters obtained from the analysis of the magnetic susceptibility, reproduce the experimental results well, including the 1/2 magnetization plateau and the zero-field excitation gap, as shown in Fig. 3 and its lower inset.
The ground state for this model is considered a singlet with the excitation gap.
We discuss the origins of these exotic quantum behaviors of the magnetizations in depth later.

\subsection{Magnetic susceptibility and magnetic specific heat in the low-temperature region}
Figures 4(a) and 4(b) show the low-temperature region of the magnetic susceptibility and the magnetic specific heat in various magnetic fields, respectively. 
The magnetic specific heats $C_{\rm{m}}$ are obtained by subtracting the lattice contribution assuming Debye's $T^3$-law as $0.0108T^3$ (J/mol K), which is applicable below about 10 K in ordinary radical compounds. 
We calculated the magnetic specific heat at zero-field by using QMC method with the obtained parameters and reproduced the experimental result with the broad peak, as shown in Fig. 4(b).
We found singular cusplike extremes in the magnetic susceptibility, as shown by the arrows in Fig. 4(a).
Such a behavior originates from the Bose-Einstein condensation of magnons, which corresponds to a phase transition to a magnetic ordered state.
The magnetic specific heats consistently exhibit field-induced sharp peaks, which are conclusive evidence for the phase transition to the ordered state, as shown in Fig. 4(b).  
We plotted these specific temperatures in a $T$-$H$ phase diagram with the double peak fields of $dM/dH$, as shown in Fig. 4(c).
A dome-like phase boundary appeared in the field-induced region, which is often observed in one-dimensional (1D) gapped spin systems. 
In the purely 1D case, the dome-like phase boundary is predicted to appear at the crossover temperature to the TLL phase as broad extremes of magnetic susceptibilities~\cite{gap2}. 
The present phase boundary is interpreted as a phase transition to the ordered state with the aid of the expected weak ferromagnetic interchain interaction.

\begin{figure}[t]
\begin{center}
\includegraphics[width=18pc]{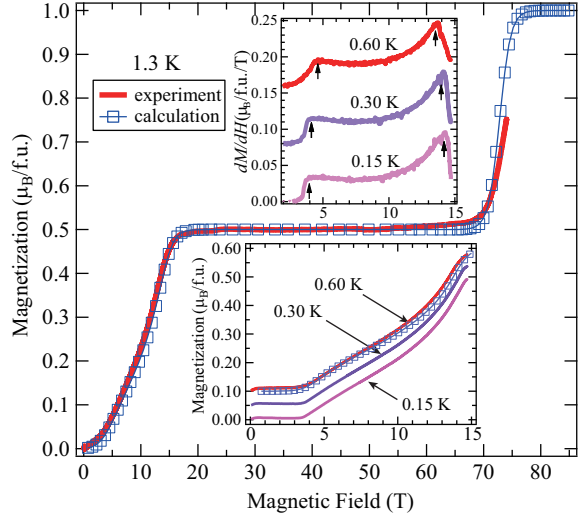}
\caption{(color online) Magnetization curve of $\beta$-2,6-Cl$_2$-V at 1.3 K in high magnetic fields. The lower inset shows those at 0.15, 0.30, and 0.60 K in low-field region, the upper inset shows their field derivative
$dM/dH$. The arrows indicate phase transition fields. For clarity, the values of the vertical axes have been shifted arbitrarily. The solid lines with open squares represent the calculated results for the $S$ = 1/2 HAFC with four-fold periodicity.}\label{f3}
\end{center}
\end{figure}

\begin{figure}[t]
\begin{center}
\includegraphics[width=20pc]{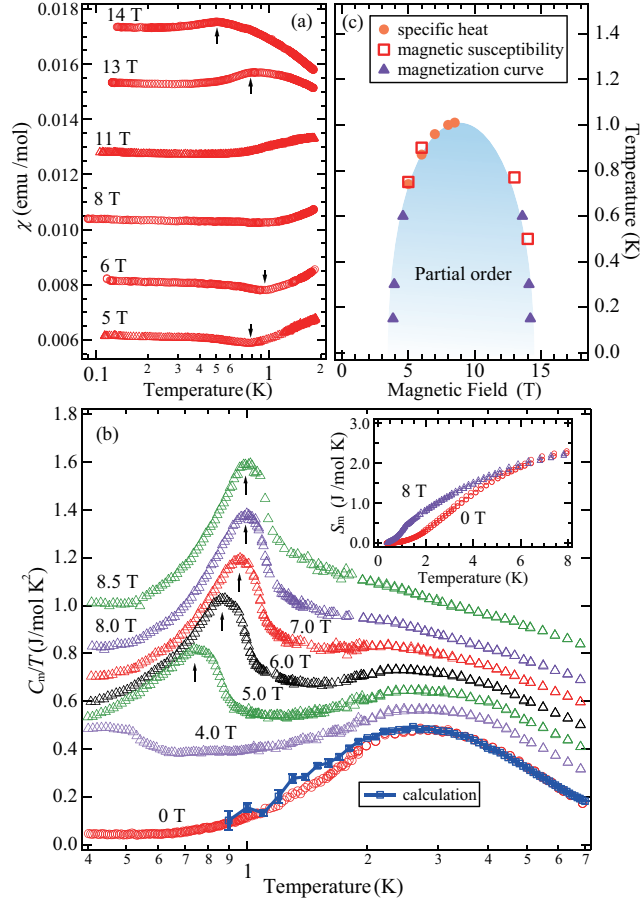}
\caption{(color online) Temperature dependence of (a) magnetic susceptibility $\chi$  and (b) $C_{\rm{m}}/T$ of $\beta$-2,6-Cl$_2$-V at various magnetic fields. For clarity, $C_{\rm{m}}/T$ for 4.0, 5.0, 6.0, 7.0, 8.0, and 8.5 T have been shifted up by 0.15, 0.25, 0.35, 0.45, 0.55, and 0.70 J/mol K$^2$, respectively. The arrows indicate phase transition temperatures. The inset shows magnetic entropy at 0 and 8.0 T. (c) Magnetic field versus temperature phase diagram of $\beta$-2,6-Cl$_2$-V. }\label{f4}
\end{center}
\end{figure}

\section{DISCUSSION}
Finally, we discuss the ground state of the present complicated spin model for the evaluated parameters.
We focus on two magnetic field regions, one below $H_{1/2}$, which is the transition field to the 1/2-plateau phase, and the other near the saturation field.
First, we consider the ground state in the former case. 
For $T \ll J_{\rm{3}}/k_{\rm{B}}$, two spins connected by the strongest AF $J_{\rm{3}}$ form a nonmagnetic singlet dimer.  
However, an effective interaction exists between the remaining spins connected by $J_{\rm{1}}$ through the triplet excited states of the singlet dimer, as described in several studies on cuprates~\cite{CuFeGeO, CuPOOH}.
The value of the effective interaction $J_{\rm{eff}}$ is given by $J_{\rm{2}}^{2}/2J_{\rm{3}}$ from the second-order perturbation treatment of the $J_{\rm{2}}$ term in the spin Hamiltonian.
Consequently, the ground state below $H_{1/2}$ can be regarded as an effective $S=1/2$ HAF alternating chain consisting of $J_{\rm{1}}/k_{\rm{B}}$ = 8.6 K and $J_{\rm{eff}}/k_{\rm{B}}$ = 7.5 K, as shown in Fig. 5(a), where the stronger $J_{\rm{1}}/k_{\rm{B}}$ form a singlet dimer with the zero-field excitation energy gap. 
The observed magnetic order at low temperatures is a long-range order of this effective alternating chain with persistent quantum spin fluctuations originating from the singlet dimmer associated with $J_{\rm{3}}$.
Correspondingly, the magnetic entropy $S_{\rm{m}}$, obtained through the integration of $C_{\rm{m}}/T$, shows that the change associated with the phase transition is only about 40$\%$ of the total entropy $S_\mathrm{m}^\mathrm{total} = R\ln 2$, as shown in the inset of Fig. 4(b).
In the 1/2-plateau phase, the energy gap of the effective alternating chain disappears, and the spins connected by $J_{\rm{1}}$ are fully polarized.
Therefore, if we express the singlet and triplet eigenstates of two spins as $|S\textgreater = 1/\sqrt{2}(|\uparrow \downarrow  \textgreater-|\downarrow \uparrow   \textgreater)$ and $|T_{\rm{1}}\textgreater =|\uparrow \uparrow \textgreater$, $|T_{\rm{0}}\textgreater = 1/\sqrt{2}(|\uparrow \downarrow  \textgreater+|\downarrow \uparrow   \textgreater)$, $|T_{\rm{-1}}\textgreater =|\downarrow \downarrow \textgreater$, respectively, we can only consider  $|T_{\rm{1}}\textgreater$ for spins connected by $J_{\rm{1}}$ and $|S\textgreater$ and $|T_{\rm{1}}\textgreater$ for spins connected by $J_{\rm{3}}$ near the saturation field at $T \ll J_{\rm{3}}/k_{\rm{B}}$.
In such a case, considering the first- and second-order perturbations of the $J_{\rm{2}}$ term in the spin Hamiltonian, the effective spin Hamiltonian $\mathcal {H}_{\rm{eff}}$ is expressed using the effective spin $\textbf{{\textit S}}_{\rm{eff}}$, whose eigenstates are $|\uparrow {\textgreater} = |T_{\rm{1}} {\textgreater} $ and $|\downarrow {\textgreater} = |S {\textgreater} $, as following XXZ model:
\begin{equation}
\mathcal {H}_{\rm{eff}} = J_{\rm{z}}{\sum^{}_{ij}}{\textit S}_{{\rm{eff}},i}^{z}{\textit S}_{{\rm{eff}},j}^{z}+J_{\rm{xy}}{\sum^{}_{ij}}({\textit S}_{{\rm{eff}},i}^{x}{\textit S}_{{\rm{eff}},j}^{x}+{\textit S}_{{\rm{eff}},i}^{y}{\textit S}_{{\rm{eff}},j}^{y})-h_{\rm{eff}}{\sum^{}_{i}}{\textit S}_{{\rm{eff}},i}^{z},
\end{equation}
where $J_{\rm{z}}=-J_{\rm{2}}^{\rm2}/8J_{\rm{3}}$, $J_{\rm{xy}}=2J_{\rm{1}}J_{\rm{2}}^{\rm2}/16J_{\rm{3}}(J_{\rm{3}}-J_{\rm{1}})$, $h_{\rm{eff}}=g{\mu _B}H-J_{\rm{3}}-J_{\rm{2}}/2-J_{\rm{2}}^{\rm2}(2J_{\rm{1}}-3J_{\rm{3}})/8J_{\rm{3}}(J_{\rm{1}}-J_{\rm{3}})$.
The evaluated values of $J_{\rm{z}}/k_{\rm{B}}$ = -1.9 K and $J_{\rm{xy}}/k_{\rm{B}}$ = 0.24 K demonstrate that the ground state can be regarded as an effective Ising ferromagnetic chain with an weak XY AF interaction, as shown in Fig. 5(b). 
There is a first-order phase transition between $|\downarrow {\textgreater}$ and $|\uparrow {\textgreater}$ at $h_{\rm{eff}}=0$, which corresponds to that between 1/2-plateau and saturated phases in the magnetization.
In this unique XXZ model, if we assume $|J_{\rm{z}}|\textless|J_{\rm{xy}}|$, an XY AF chain model becomes effective, and a TLL consisting of the effective spins expected to be realized.  

\begin{figure}[t]
\begin{center}
\includegraphics[width=18pc]{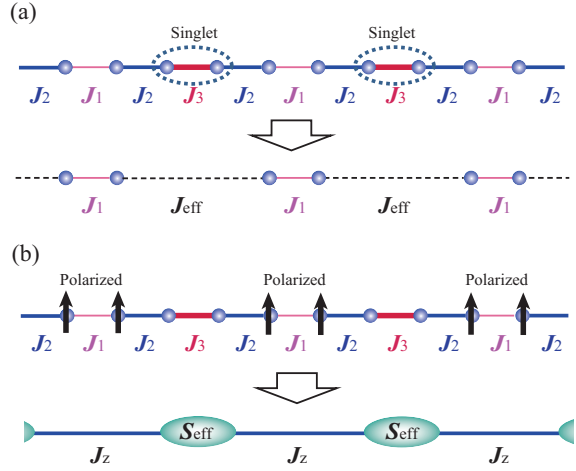}
\caption{(color online) Effective magnetic models in $\beta$-2,6-Cl$_2$-V at $T \ll J_{\rm{3}}/k_{\rm{B}}$. (a) The $S$ = 1/2 HAF alternating chain formed by $J_{\rm{1}}$ and $J_{\rm{eff}}$ below $H_{1/2}$. Broken ellipses indicate singlet pairs connected by $J_{\rm{3}}$. (b) The Ising ferromagnetic chain formed by $J_{\rm{z}}$ with a weak XY AF interaction in terms of the effective spin $\textbf{{\textit S}}_{\rm{eff}}$ near the saturation field. The arrows indicate fully polarized spins connected by $J_{\rm{1}}$.}\label{f3}
\end{center}
\end{figure}

\section{SUMMARY}
We have succeeded in synthesizing a verdazyl radical crystal of $\beta$-2,6-Cl$_2$-V.
The $ab$ $initio$ MO calculation indicated the formation of an $S=1/2$ Heisenberg AF chain with four-fold magnetic periodicity consisting of three types of exchange interactions.
We have successfully explained the magnetic and thermodynamic properties based on the expected spin model by using the QMC method.
In the low magnetic field regions, one of the three interactions and the effective interaction form the $S=1/2$ Heisenberg AF alternating chain.
The observed field-induced ordered state is a long-range order of this effective chain with persistent quantum spin fluctuations.
Near the saturation field, contrary to the fundamental Heisenberg AF chain, the Ising ferromagnetic chain with a weak XY AF interaction becomes effective.
We demonstrated that verdazyl radical could form unconventional spin model with interesting quantum behavior.
These results provides a new way to synthesize magnetic materials with a variety of quantum spin systems resulting in further our understanding of quantum effect in magnetic materials.

\begin{acknowledgments}
We thank T. Shimokawa, T. Tonegawa, and S. Todo for the valuable discussions. This research was partly supported by KAKENHI (Nos. 24740241, 24540347, and 24340075)DA part of this work was performed under the interuniversity cooperative research program of the joint-research program of ISSP, the University of Tokyo. This work was partially supported by the Strategic Programs for Innovative Research (SPIRE), MEXT, and the Computational Materials Science Initiative (CMSI), Japan. Some computations were performed using the facilities of the Supercomputer Center, ISSP, The University of Tokyo. Our QMC calculations were carried out using the ALPS application~\cite{ALPS}.
\end{acknowledgments}


\end{document}